\documentstyle[prl,eqsecnum,aps]{revtex}

\begin{document}
\author{ Jian-Qi Shen \footnote{E-mail address: jqshen@coer.zju.edu.cn}}
\address{Zhejiang Institute of Modern Physics and Department of Physics,\\
Zhejiang University, Hangzhou 310027, P. R. China}
\date{\today}
\title{Gravitational Effects Associated With Gravitomagnetic Fields}
\maketitle

\begin{abstract}
Several relativistic quantum gravitational effects such as spin-rotation
coupling, gravitomagnetic charge and gravitational Meissner effect are
investigated in the present letter. The field equation of gravitomagnetic
matter is suggested and a static spherically symmetric solution of this
equation is offered. With foreseeable improvements in detecting and
measuring technology, it is possible for us to investigate quantum mechanics
in weak-gravitational fields. The potential implications of these
gravitational effects (or phenomena) to some problems are briefly discussed.
\end{abstract}

\pacs{PACS:}

It is readily verified that the field equation of general relativity in
low-motion weak-field approximation is somewhat analogous to Maxwell$^{,}$s
equation of electromagnetic field. The most outstanding point is that the
former field (gravitational field) possesses both the gravitoelectric
potential written as $\frac{g_{00}-1}{2}$ and the gravitomagnetic potentials
as $\vec{A}=(g_{01},g_{02},g_{03}),$ and the corresponding gravitomagnetic
field strength is of the form $\vec{B}=-\frac{1}{2}\nabla \times \vec{A}$.
This similarity leads us to consider the following interesting gravitational
analogues of electromagnetic phenomena: (1) In electrodynamics a charged
particle is acted upon by the Lorentz magnetic force, in the similar
fashion, a particle is also acted upon by the gravitational Lorentz force in
weak-gravity theory \cite{Shen}. According to the principle of equivalence,
further analysis shows that in the non-inertial rotating reference frame,
this gravitational Lorentz force is just the fictitious Coriolis force; (2)
There exists Aharonov-Bohm effect in electrodynamics\cite{Bohm},
accordingly, the so-called gravitational Aharonov-Bohm effect, the
gravitational analogue of Aharonov-Bohm effect also exists in the theory of
gravitation, which is now termed Aharonov-Carmi effect\cite{Aharonov}; (3) A
particle with intrinsic spin possesses a gravitomagnetic moment of such
magnitude that it equals the spin of this particle. The interaction of
spinning gravitomagnetic moment with the gravitomagnetic field is called
spin-gravity coupling, which is similar to the interaction between spinning
magnetic moment and magnetic field in electrodynamics. For the present, it
is possible to investigate quantum mechanics in weak-gravitational fields,
with the development of detecting and measuring technology, particularly
laser-interferometer technology, low-temperature technology, electronic
technology and so on. These investigations enable physicists to test
validity or universality of fundamental laws and principles of general
relativity. In this letter, the author proposes and discusses several
physically interesting effects associated with gravitomagnetic fields such
as (1) the geometric phase factor in the time-dependent spin-rotation
coupling, (2) gravitomagnetic charge (dual mass), (3) the motion of photon
in a gravitomagnetic field and some other effects, for instance,
gravitational Meissner effect which may lead to significant consequences in
astrophysics and cosmology.

In the rotating reference frame, a particle was acted on by the inertial
centrifugal force and Coriolis force $2m\vec{v}\times \vec{\omega}$, which
are respectively in analogy with the electric force and Lorentz magnetic
force $q\vec{v}\times \vec{B}$ in electrodynamics, where $\vec{\omega}$ and $%
\vec{B}$ denote rotating the frequency of rotating frame and the magnetic
field strength, respectively. It follows from this comparison between
Coriolis force and Lorentz force that $\vec{\omega}$ is believed to be
regarded as the gravitomagnetic field in the rotating frame. One can easily
obtain the Lagrangian, $L$ and Hamiltonian, $H$ of the particle in the
rotating frame \cite{Li}, i.e.,
\begin{eqnarray}
L &=&\frac{1}{2}mv^{2}-m\phi +2m\vec{v}\cdot \vec{a},  \nonumber \\
H &=&\frac{1}{2m}(\vec{p}-2m\vec{a})^{2}+m\phi ,  \eqnum{1}
\end{eqnarray}
where the canonical momentum $\vec{p}=m\vec{v}+2m\vec{a},$ and
gravitomagnetic vector potential $\vec{a}$ is so defined that $\nabla \times
\vec{a}=\vec{\omega}.$ Aharonov and Carmi showed that a geometric quantum
phase factor which is given as (in the unit $\hbar =1$)
\begin{equation}
\Delta \theta =2m\oint_{abcda}\vec{a}\cdot d\vec{l}=2m\vec{\omega}\cdot \vec{%
A}  \eqnum{2}  \label{eq2}
\end{equation}
would arise in the wave function of a particle moving along a closed path $%
abcda$ surrounding a gravitomagnetic flux, $\vec{\omega},$ where $\vec{A}$
represents the area vector that is surrounded by the path $abcda$.
Overhauser, Colella \cite{Overhauser}, Werner and Standenmann et al.\cite
{Werner} have proved the existence of the Aharonov-Carmi effect by means of
the neutron-gravity interferometry experiment. Further calculation shows
that when the Kerr metric is transformed from the fixing reference frame to
the rotating frame, the gravitomagnetic potentials can be written as (in the
spherical coordinate system)
\begin{equation}
a_{\varphi }=\frac{aGM\sin \theta }{cr^{2}}+\omega r\sin \theta ,\quad
a_{r}=0,\quad a_{\theta }=0.  \eqnum{3}  \label{eq3}
\end{equation}
It follows that the first term $\frac{aGM\sin \theta }{cr^{2}}$ on the right
handed side of Eq. (\ref{eq3}) is exactly analogous to the magnetic
potential $\frac{\mu _{0}}{4\pi }\frac{ea}{r^{2}}\sin \theta $ of the
rotating charged spherical shell in electrodynamics. Then we can calculate
the exterior gravitomagnetic strength of the rotating gravitating
spherically symmetric body, and the result is $\vec{B}_{g}=\frac{2G}{c}(%
\frac{\vec{a}}{r^{3}}-\frac{3(\vec{a}\cdot \vec{r})\vec{r}}{r^{5}})\cite
{Ahmedov}.$ The second term of $a_{\varphi }$ yields the gravitomagnetic
field, $\vec{\omega}$ which is independent of the Newtonian gravitational
constant $G.$

In the following we will derive the Hamiltonian of spin-rotation coupling by
investigate the Dirac equation with spin connection

\begin{equation}
\lbrack i\gamma ^{\mu }(\partial _{\mu }-\frac{i}{4}\sigma ^{\lambda \tau
}\omega _{\lambda \tau \mu })-mc]\psi =0  \eqnum{4}  \label{eq9}
\end{equation}
with $\sigma ^{\lambda \tau }=\frac{i}{2}(\gamma ^{\lambda }\gamma ^{\tau
}-\gamma ^{\tau }\gamma ^{\lambda }).$ In the rotating frame, we have the
following form of the line element of spacetime

\begin{equation}
ds^{2}=(1-\frac{\omega ^{2}}{c^{2}}\vec{x}\cdot \vec{x})c^{2}dt^{2}-d\vec{x}%
\cdot d\vec{x}-2(\vec{\omega}\times \vec{x})\cdot d\vec{x}dt  \eqnum{5}
\label{eq11}
\end{equation}
by ignoring the gravitational effect associated with the gravitational
constant $G$ and utilizing the weak-field low-motion approximation. Then
further calculation yields the following connections\cite{Hehl}

\begin{eqnarray}
\omega _{\lambda \tau 0} &=&-\epsilon _{\lambda \tau \eta }\frac{\omega
^{\eta }}{c},\quad \omega _{0\tau 0}=-\omega _{\tau 00}=0,\quad \quad
\nonumber \\
\omega _{\lambda \tau \mu } &=&0\quad (\mu =1,2,3)  \eqnum{6}  \label{eq10}
\end{eqnarray}
with $\epsilon _{\lambda \tau \eta }$ being three-dimensional Levi-Civita
tensor. By making use of Eq. (\ref{eq9}), Eq. (\ref{eq11}) and Eq. (\ref
{eq10}), one can arrive at the following Dirac equation

\begin{equation}
i\frac{\partial }{\partial t}\psi =H\psi  \eqnum{7}  \label{eq12}
\end{equation}
with

\begin{equation}
H=\beta mc^{2}+c\cdot \vec{p}+\vec{\omega}\cdot \vec{L}+\vec{\omega}\cdot
\vec{S}.  \eqnum{8}  \label{EQ13}
\end{equation}
We thus obtain the Hamiltonian of spin-rotation coupling

\begin{equation}
H_{s-r}=\vec{\omega}\cdot \vec{S}  \eqnum{9}  \label{eq14}
\end{equation}
which is consistent with Mashhoon$^{,}$s result which is obtained by
analyzing the Doppler$^{,}$s effect of wavelight in the rotating frame with
respect to the fixing frame \cite{Mashhoon1,Mashhoon2}.

It is well known that geometric phase appears in systems whose Hamiltonian
is time-dependent or possessing evolution parameters. Apparently, the
geometric phase in the Aharonov-Carmi effect results from the presence of
the evolution parameter in the Hamiltonian. Here we suggest another
geometric effect that exist in the spin-rotation coupling where the rotating
frequency, $\vec{\omega}$ of the rotating frame is time-dependent. By making
use of the Lewis-Riesenfeld invariant theory\cite{Lewis} and the
invariant-related unitary transformation formulation\cite{Gao}, one can
arrive at \
\begin{equation}
\Delta \theta =\pm \frac{1}{2}\int_{0}^{t}\dot{\gamma}(t^{^{\prime
}})[1-\cos \lambda (t^{^{\prime }})]dt^{^{\prime }}  \eqnum{10}  \label{eq15}
\end{equation}
which is the expression for the geometric phase of a particle with spin-$%
\frac{1}{2},$ e.g., a neutron in the time-dependent rotating frame, where $%
\gamma $ and $\lambda $ stand for the angle displacements in the parameter
space of the invariant whose eigenvalue is time-independent, and dot denotes
the rate of change of $\gamma $ with respect to $t$. For the case of
adiabatic process where $\lambda $ does not explicitly involve time, $t,$
Eq. (\ref{eq15}) is reduced to
\begin{equation}
\Delta \theta =\pm \frac{1}{2}\cdot 2\pi (1-\cos \lambda )  \eqnum{11}
\label{eq16}
\end{equation}
in one cycle over the parameter space. It follows from Eq. (\ref{eq16}) that
$2\pi (1-\cos \lambda )$ is the expression for the solid angle which
presents the geometric properties of time evolution of this spin-rotation
coupling system. Differing from the dynamical phase which is related to the
energy, frequency or velocity of a particle or a quantum system, geometric
phase is dependent only on the geometric nature of the pathway along which
the system evolves. This property reflects the global and topological
properties of evolution of the quantum systems\cite{Berry,Simon}.

Another topological property of gravitation is the gravitational analogue of
Dirac monopole in electrodynamics. By using variational principle, we can
obtain the following equation
\begin{equation}
\delta \int_{\Omega }\sqrt{-g}\tilde{R}d\Omega =\int_{\Omega }\sqrt{-g}%
\tilde{G}_{\mu \nu }\delta g^{\mu \nu }d\Omega   \eqnum{12}  \label{eq17}
\end{equation}
with
\begin{equation}
\tilde{R}=g^{\sigma \tau }\tilde{R}_{\sigma \tau },\quad \tilde{R}_{\sigma
\tau }=g^{\mu \delta }(\epsilon _{\mu \upsilon }^{\quad \ \alpha \beta
}R_{\gamma \delta \alpha \beta }+\epsilon _{\gamma \delta }^{\quad \ \alpha
\beta }R_{\mu \upsilon \alpha \beta })  \eqnum{13}  \label{eq18}
\end{equation}
and
\begin{equation}
\tilde{G}_{\mu \nu }=\epsilon _{\mu }^{\quad \lambda \sigma \tau }R_{\nu
\lambda \sigma \tau }-\epsilon _{\nu }^{\quad \lambda \sigma \tau }R_{\mu
\lambda \sigma \tau },  \eqnum{14}  \label{eq19}
\end{equation}
where $\epsilon _{\mu }^{\quad \lambda \sigma \tau }$ is the completely
antisymmetric Levi-Civita tensor, and the scalar $\tilde{R}$ is assumed to
be the Lagrangian density of the interaction of metric fields with
gravitomagnetic matter. Since $\tilde{G}_{\mu \nu }$ is an antisymmetric
tensor, we construct the following antisymmetric tensor for the Fermi field
\begin{equation}
K_{\mu \nu }=i\bar{\psi}(\gamma _{\mu }\partial _{\nu }-\gamma _{\upsilon
}\partial _{\mu })\psi ,\quad H_{\mu \nu }=\epsilon _{\mu \nu }^{\quad
\alpha \beta }K_{\alpha \beta }  \eqnum{15}  \label{eq20}
\end{equation}
and regard them as the source tensors in the field equation of
gravitomagnetic charge, where $\gamma _{\mu }^{\quad ^{,}}$s are general
Dirac matrices with respect to $x^{\mu }$ and satisfy $\gamma _{\mu }\gamma
_{v}+\gamma _{\nu }\gamma _{\mu }=2g_{\mu \nu }.$ Then the field equation
governing the gravitational distributions of gravitomagnetic charge may be
given as follows
\begin{equation}
\tilde{G}_{\mu \nu }=\kappa _{1}K_{\mu \nu }+\kappa _{2}H_{\mu \nu }
\eqnum{16}  \label{eq21}
\end{equation}
with $\kappa _{1},\kappa _{2}$ being the coupling coefficients between
gravitomagnetic matter and gravity. It should be noted that $\tilde{G}_{\mu
\nu }\equiv 0$ in the absence of gravitomagnetic matter since no
singularities associated with topological charge exist in the metric
functions and therefore Ricci identity still holds. However, once the metric
functions possess non-analytic properties in the presence of gravitomagnetic
matter (should such exist), $\tilde{G}_{\mu \nu }$ is no longer vanishing
due to the violation of Ricci identity. Additionally, further investigation
shows that the {\sl cosmological term }of Fermi field in Eq. (\ref{eq21})
can be written as the linear combination of the antisymmetric tensors $i\bar{%
\psi}(\gamma _{\mu }\gamma _{\nu }-\gamma _{\upsilon }\gamma _{\mu })\psi $
and $i\epsilon _{\mu \nu }^{\quad \alpha \beta }\bar{\psi}(\gamma _{\alpha
}\gamma _{\beta }-\gamma _{\beta }\gamma _{\alpha })\psi $.

It is verified that Eq. (\ref{eq21}) in the low-motion weak-field
approximation is of the form
\begin{equation}
\nabla \times (\nabla g^{00}-\frac{\partial }{\partial x^{0}}\vec{g})=-\frac{%
\partial }{\partial x^{0}}(\nabla \times \vec{g})+\vec{S}  \eqnum{17}
\label{eq22}
\end{equation}
with gravitomagnetic vector potentials $\vec{g}=(g^{01},g^{02},g^{03}),$ and
source-current density, $\vec{S}$ of gravitomagnetic charge being $\kappa
_{1}K^{0i}+\kappa _{2}H^{0i}$ $(i=1,2,3).$ It is apparently seen that Eq. (%
\ref{eq22}) is exactly analogous to the Faraday$^{^{,}}$s law of
electromagnetic induction in the presence of current density of magnetic
monopole in electrodynamics. This, therefore, implies that Eq. (\ref{eq21})
is indeed the field equation of gravitation of gravitomagnetic matter.

Assume that a point-like gravitomagnetic charge is fixed at the origin of
the spherical coordinate system. A static spherically symmetric solution is
exactly obtained by the author and is given (via linear element) as follows
\begin{eqnarray}
ds^{2} &=&(dx^{0})^{2}-dr^{2}-r^{2}(d\theta ^{2}+\sin ^{2}\theta d\varphi
^{2})+2g_{0\varphi }dx^{0}d\varphi ,  \nonumber \\
2g_{0\varphi }dx^{0}d\varphi  &=&\mp \frac{2c}{4\pi }\cdot \frac{1\pm \cos
\theta }{r\sin \theta }\cdot r\sin \theta dx^{0}d\varphi ,  \eqnum{18}
\label{eq23}
\end{eqnarray}
where $c$ is defined to be determined by the metric functions of the origin
of the spherical coordinate system, i.e., $c=-M(\frac{\sqrt{-g}}{g^{00}}%
)_{origin},$ and $M$ is the parameter associated with gravitomagnetic charge
and coupling coefficients. Further calculation yields
\begin{equation}
\left( g^{\mu \nu }\right) =\left(
\begin{array}{cccc}
\frac{r^{2}\sin ^{2}\theta }{r^{2}\sin ^{2}\theta +g_{0\varphi }^{2}} & 0 & 0
& \frac{g_{0\varphi }}{r^{2}\sin ^{2}\theta +g_{0\varphi }^{2}} \\
0 & -1 & 0 & 0 \\
0 & 0 & -r^{-2} & 0 \\
\frac{g_{0\varphi }}{r^{2}\sin ^{2}\theta +g_{0\varphi }^{2}} & 0 & 0 &
\frac{-1}{r^{2}\sin ^{2}\theta +g_{0\varphi }^{2}}
\end{array}
\right)   \eqnum{19}  \label{eq24}
\end{equation}
which is the inverse matrix of the metric $\left( g_{\mu \nu }\right) $ and
we thus obtain the contravariant metric $g^{\mu \nu }.$

From what has been discussed above, similar to the magnetic charge in
electrodynamics, gravitomagnetic charge is a kind of topological charge
which is the duality of mass of matter. In this sense, gravitomagnetic
charge is also called dual mass. From the point of view of the classical
field equation, matter may be classified into two categories:
gravitomagnetic matter and gravitoelectric matter, of which the field
equation of the latter is Einstein$^{,}$s equation of general relativity.
Due to their different gravitational features, the concept of mass is of no
significance for the gravitomagnetic matter. It is also of interest to
investigate the relativistic dynamics of dual mass.

It is worthwhile to consider the motion of photon in gravitomagnetic fields.
Take into account a weak gravitomagnetic fields where the adiabatic
approximation can be applicable to the motion of a photon. Then a conclusion
can be drawn that the eigenvalue of the helicity $\frac{\vec{k}(t)}{k}\cdot
\vec{J}$ of the photon is conserved in motion and its helicity operator $%
\frac{\vec{k}(t)}{k}\cdot \vec{J}$ is an invariant in terms of the invariant
equation in Lewis-Riesenfeld theory\cite{Lewis}
\begin{equation}
\frac{\partial I(t)}{\partial t}+\frac{1}{i}[I(t),H(t)]=0,  \eqnum{20}
\label{eq25}
\end{equation}
where the invariant $I(t)=\frac{\vec{k}(t)}{k}\cdot \vec{J}.$ From Eq. (\ref
{eq25}), simple calculation yields
\begin{equation}
H(t)=\frac{\vec{k}\times \frac{d}{dt}\vec{k}}{k^{2}}\cdot \vec{J}  \eqnum{21}
\label{eq26}
\end{equation}
which is considered an effective Hamiltonian governing the motion of photon
in gravitomagnetic fields. Hence, the equation of motion of a photon in
gravitomagnetic fields is written
\begin{equation}
\frac{\vec{k}\times \frac{d}{dt}\vec{k}}{k^{2}}=\vec{B}_{g},  \eqnum{22}
\label{eq27}
\end{equation}
where the gravitomagnetic field strength $\vec{B}_{g}$ is determined by
field equations of gravitation such as Eq. (\ref{eq21}) and Einstein$^{,}$s
equation of general relativity. It follows from the geodesic equation under
weak-field approximation that the acceleration due to gravitational Lorentz
force is
\begin{equation}
\frac{1}{k}\frac{d}{dt}\vec{k}=-\frac{\vec{k}}{k}\times (\nabla \times \vec{g%
})  \eqnum{23}  \label{eq28}
\end{equation}
with $\frac{\vec{k}}{k}$ being the velocity vector of the photon, and the
gravitomagnetic field strength is $\vec{B}_{g}=\nabla \times \vec{g},$ where
$\vec{g}=(g^{01},g^{02},g^{03}).$ Substitution of Eq. (\ref{eq27}) into Eq. (%
\ref{eq28}) yields
\begin{equation}
\frac{1}{k}\frac{d}{dt}\vec{k}=-\frac{\vec{k}}{k}\times \left( \frac{\vec{k}%
\times \frac{d}{dt}\vec{k}}{k^{2}}\right) .  \eqnum{24}  \label{eq29}
\end{equation}
Since
\begin{equation}
\vec{k}\cdot \vec{k}=k^{2},\quad \vec{k}\cdot \frac{d}{dt}\vec{k}=0,
\eqnum{25}  \label{eq30}
\end{equation}
Eq. (\ref{eq29}) is an identity. This, therefore, implies that Eq. (\ref
{eq27}) is the equation of motion of a photon in gravitomagnetic field.

Note that when observed from the rotating reference frame, the
gravitomagnetic field $\nabla \times \vec{g}$ contains the rotating
frequency, $\vec{\omega}$ (see, for instance, Eq. (\ref{eq3}))$.$ For the
reason of that $\vec{\omega}$ is an effective gravitomagnetic field that is
independent of Newtonian gravitational constant, the fact that $\vec{\omega}$
is involved in $\nabla \times \vec{g}$ results in both the Aharonov-Carmi
effect and spin-rotation coupling.

What we have discussed above are some geometric or topological features
which relate to gravitomagnetic fields. In what follows the author will
briefly discuss several other physically interesting problems. They are
illustrated as follows:

(1) The geometric phase expressed by Eq. (\ref{eq15}) is believed to be a
potential application to the Earth$^{^{,}}$s time-dependent rotating
frequency. The variation of the Earth$^{,}$s rotating frequency may be
caused by the motion of interior matter, tidal force, and the motion of
atmosphere as well. Once we have information concerning the Earth$^{,}$s
rotating frequency, it is possible to investigate the motion of matter on
the Earth. By measuring the geometric phase difference of spin polarized
vertically down and up in the neutron-gravity interferometry experiment,
utilization of Eq. (\ref{eq15}) may enable us to obtain the information
concerning the variation of the Earth$^{^{,}}$s rotation.

(2) The gravitational analogue of Meissner effect in superconductivity is
gravitational Meissner effect. Due to the conservation of momentum,
mass-current density is conserved around the scattering of particles in
perfect fluid, which is analogous to the superconductivity of
superconducting electrons in superconductors cooled below $T_{c}.$ Since
gravitational field equation under linear approximation is similar to the
Maxwell equation in electrodynamics, one can predict that gravitational
Meissner effect arises in perfect fluid. The author think that the
investigation of both the effect of gravitomagnetic matter and gravitational
Meissner effect may provide us with a valuable insight into the problem of
cosmological constant and vacuum gravity\cite{Weinberg}. Some related topics
such as gravitational Hall effect and gravitational magnetohydrodynamics may
be considered for further investigation.

(3) Taking the effects of gravitomagnetic charge into consideration is
believed to be of essential significance in resolving some problems and
paradoxes. An illustrative example that has been briefly discussed is its
application to the problem of cosmological constant. Additionally, in 1998,
Anderson et al reported that, by ruling out a number of potential causes,
radio metric data from the Pioneer 10/11, Galileo and Ulysses spacecraft
indicate an apparent anomalous, constant, acceleration acting on the
spacecraft with a magnitude $\sim 8.5\times $ $10^{-8}$cm/s$^{2}$ directed
towards the Sun\cite{Anderson}. Is it the effects of dark matter or a
modification of gravity? Unfortunately, neither easily works. By taking the
cosmic mass, $M=10^{53}$ kg and cosmic scale, $R=10^{26}$ m, calculation
shows that this acceleration is just equal to the value of field strength on
the cosmic boundary due to the total cosmic mass. This fact leads us to
consider a theoretical mechanism to interpret this anomalous phenomenon. The
author favor that the gravitational Meissner effect may serve as a possible
interpretation. Here we give a rough analysis which contains only the most
important features rather than the precise details of this theoretical
explanation. Once gravitomagnetic matter exists in the universe, parallel to
London$^{,}$s electrodynamics of superconductivity, it shows that
gravitational field may give rise to an {\sl effective mass }$m_{g}=\frac{%
\hbar }{c^{2}}\sqrt{8\pi G\rho _{m}}$ due to the self-induced charge current
\cite{Hou}, where $\rho _{m}$ is the mass density of the universe. Then one
can obtain that $\frac{\hbar }{m_{g}c}\simeq 10^{26}$m which approximately
equals $R$, where the mass density of the universe is taken to be $\rho
_{m}=0.3\rho _{c}$ with $\rho _{c}\simeq 2\times 10^{-26}$kg/m$^{3}$ being
the critical mass density. An added constant acceleration, $a$ may result
from the Yukawa potential and can be written as
\begin{equation}
a=\frac{GM}{2}\left( \frac{m_{g}c}{\hbar }\right) ^{2}\simeq \frac{GM}{R^{2}}%
.  \eqnum{26}  \label{eq31}
\end{equation}
Note, however, that it is an acceleration of repulsive force directed,
roughly speaking, from the centre of the universe. By analyzing the NASA$^{,}
$s Viking ranging data, Anderson, Laing, Lau et al concluded that the
anomalous acceleration does not act on the body of large mass such as the
Earth and Mars. If gravitational Meissner effect only affected the
gravitating body of large mass or large scale rather than spacecraft
(perhaps the reason lies in that small-mass flow cannot serve as
self-induced charge current, which deserves to be further investigated),
then seen from the Sun or Earth, there exists an added attractive force
acting on the spacecraft. This added force give rise to an anomalous,
constant, acceleration directed towards the Sun or Earth.

In summary, in the present paper the author investigates the geometric phase
that results from the time-dependent spin-rotation coupling, and the field
equation of gravitomagnetic matter as well as its static spherically
symmetric solution. Differing from the symmetric property (with respect to
the indices of tensors) of gravitational field equation of gravitoelectric
matter, the field equation of gravitomagnetic matter possesses the
antisymmetric property. This, therefore, implies that the number of the
non-analytic metric functions is no more than 6. Although we have no
observational evidences for the existence of gravitomagnetic charge, it is
still of essential significance to investigate the gravity theory of the
topological dual mass.

Some physically interesting problems associated with gravitomagnetic fields
are proposed, of which the most intereting investigation is the potential
solution to the anomalous acceleration acting on Pioneer spacecrafts by
means of the mechanism of gravitational Meissner effect. The theoretic
resolution of this problem is not very definite at present, for it cannot
account for the fact that the gravitational Meissner effect does not
explicitly affect the body of small mass. This curiosity deserves further
consideration.

Since with foreseeable improvements in detecting and measuring technology,
it is possible for us to investigate quantum mechanics in weak-gravitational
fields, the above effects and phenomena deserve further detailed
investigations. Work in this direction is under consideration and will be
published elsewhere.

Acknowledgments This project is supported by the National Natural Science
Foundation of China under the project No.$30000034$.

\end{document}